\documentclass[10pt,twocolumn,presuperscriptaddress,showpacs]{revtex4-1}

\usepackage{graphicx}
\usepackage{epstopdf}
\usepackage{amsmath,amssymb}
\usepackage{color}

\begin{document}

\title{Non-scaling displacement distributions as may be seen in fluorescence correlation spectroscopy}
\author{S.M.J. Khadem}
\affiliation{Institute of Physics, Humboldt University Berlin, Newtonstr. 15, D-12489 Berlin, Germany}
\affiliation{School of Analytical Sciences Adlershof (SALSA) Albert-Einstein-Str. 5-9, D-12489 Berlin, Germany}
%
\author{I.M. Sokolov}
\affiliation{Institute of Physics, Humboldt University Berlin, Newtonstr. 15, D-12489 Berlin, Germany}
\affiliation{School of Analytical Sciences  Adlershof (SALSA) Albert-Einstein-Str. 5-9, D-12489 Berlin, Germany}

\begin{abstract}
  A continuous time random walk (CTRW) model with waiting times following the L\'evy-stable distribution with exponential cut-off in equilibrium is a simple
  theoretical model giving rise to normal, yet non-Gaussian diffusion. The distribution of the particle's displacements is explicitly time-dependent
  and does not scale. Since fluorescent correlation spectroscopy (FCS) is often used to investigate diffusion processes, we discuss the influence of this
  lack of scaling on the possible outcome of the FCS measurements and calculate the FCS autocorrelation curves for such equilibrated CTRWs. 
  The results show that although the deviations from Gaussian behavior may be detected when
analyzing the short- and long-time asymptotic behavior of the corresponding curves,
their bodies are still perfectly fitted by the fit forms used for normal diffusion. The diffusion coefficients obtained from the fits may however differ 
considerably from the true tracer diffusion coefficients as describing the time dependence of the mean squared displacement. 
\end{abstract}


\maketitle

\section{Introduction}

Diffusion is an ubiquitous transport process stemming from the particles' thermal motion. 
Normal diffusion is the dominant form of diffusion, and  has been explained in the works by 
Einstein \cite{eins}, Smoluchowski \cite{smol} and Langevin \cite{lang}.
While the Einstein's and Smoluchowski's approaches were essentially based on the discussion of a random walk scheme, 
Langevin's one relied on the stochastic differential equation with a random force acting on a particle. These approaches 
are widely in use until now. 

The probability density function  (PDF) of particle's' displacement in normal diffusion follows a
Gaussian distribution which obeys the standard diffusion equation (second Fick's law). 
The mean squared displacement $\delta r^2$ of the particles in such diffusion grows linearly
in time: $\delta r^2 \propto t$. 

Normal diffusion is observed for passive particles moving in homogeneous fluid environments. 
Diffusion behaviour in complex environments may show  a different time dependence, $\delta r^2 \propto t^\alpha$
with $\alpha \neq 1$, termed anomalous diffusion;
$\alpha <1$ correspond to subdiffusion, and $\alpha >1$ to the superdiffusion. 

Different physical models have been proposed to explain diffusion anomalies; they are based on different physical 
assumptions and their description relies of very different mathematical tools, being the generalizations of the 
Einstein's and Langevin's approaches \cite{sokolov2012models}. 

Recently, in some  simulations and in experiments using single particle tracking (SPT) \cite{spt}, normal diffusion behavior
$\delta r^2 \propto t$ has been observed together with non-Gaussian PDFs. 
Diffusion of colloidal beads along linear phospholipid bilayer tubes and in biofilament network \cite{1}, enhanced tracer diffusion in suspensions of microorganisms \cite{2}, 
liposomes in nematic solutions of aligned F-actin filaments \cite{3}, nanoparticles in hard-sphere colloidal suspensions \cite{4}, colloidal hard discs in solid and hexatic phases \cite{5} 
are examples showing such property. In some of these situations the deviation from Gaussian form is significant only at short times, while at longer times the PDF tends to the 
Gaussian form typical for the normal diffusion. This behavior can be phenomenologically captured by a theoretical model of `Diffusing Diffusivity` \cite{bia3}.
Since the form of the PDF changes in course of the time, it does not scale.

One of the methods routinely used in investigation of diffusive behaviour in complex systems (along with SPT) is the fluorescence correlation
spectroscopy (FCS) \cite{elson}. It is a method of choice in investigation of diffusion and anomalous diffusion in soft matter and in biological systems. 
This indeed can detect the anomalies \cite{weiss2003}, and also the deviations from the Gaussian form of the PDF, as it was shown in \cite{us}.
The standard implementations of the methods assume however the scaling of the displacements' PDF as a whole, as it is the case for the normal and
fractional Brownian motion, and in percolation. Therefore it is necessary to understand, what could be seen experimentally in the cases 
when the scaling is absent. The best way to proceed here is to consider first a simple model which to a large extent can be treated analytically. 
As such a model we chose the continuous time random walk (CTRW) scheme \cite{ctrw} with waiting time distribution given by a truncated power law.  
   
CTRW is a versatile model generalizing the Einstein's description of normal diffusion to some of the anomalous cases.
Random walker described by this model performs jumps between different positions with waiting time between the jumps chosen from the 
waiting time distribution with a probability density function $\psi(t)$. Choosing this PDF to follow a power law  $\psi(t) \propto 1/t^{1+\alpha}$ 
not possessing the first moment (i.e. with $0< \alpha < 1$) leads to subdiffusion. 
Using a CTRW model with power-law waiting time density in \cite{scher1975anomalous} lead to a successful explanation of the transient photocurrent traces in the 
time-of-flight experiments in strongly disordered semiconductors. In such experiments, the charge carriers are created by a light flash at the 
beginning of the measurement, and the ordinary CTRW process (with the beginning of observation coinciding with the first step of a walker) is the relevant model. 
On the other hand, in many experiments, especially in biological systems, the experiment is performed on a pre-existing system, or the techniques involve 
long data acquisition times. In this case the process at the beginning of observation is already strongly aged or even fully equilibrated. 
In this case the CTRW with a power-law waiting time distribution shows extremely slow diffusion, if any diffusion at all, while
a CTRW with any waiting time PDF possessing the first moment (e.g. with a truncated power-law waiting time distribution) generically leads to normal 
diffusion in the entire time domain \cite{tunaley1975}. However in this situation, the PDF of the particles' displacements PDF may still be strongly non-Gaussian 
for time lages shorter than the truncation time, as is known since long in mathematically similar econometric models \cite{clark1973subordinated}. 
Such  models can be proper candidates to explain the observed experimental results and may be useful to gain more information on systems showing such behaviour. 

In what follows we discuss the corresponding model using an example of truncated one-sided L\'evy stable PDF 
of waiting times, and concentrate mostly on the application of our results to fluorescence correlation spectroscopy. 
The FCS under ordinary CTRW process (as described by a fractional diffusion equation)
was discussed in detail in Ref. \cite{klafter}. This process leads to anomalous diffusion in the whole time range and shows such exotic properties as aging and ergodicity breaking. 
In what follows, we discuss the same problem for a fully equilibrated CTRW model with a waiting time distribution following one-sided L\'evy PDF 
with an exponential cut-off at long times. The model shows normal diffusion in the whole time domain, but the PDF of particles' displacements shows a crossover 
from a non-Gaussian form at shorter times to a Gaussian asymptotic long-time behaviour, which is clearly seen in the behaviour of the fourth moment of the corresponding
distribution. Then we calculate the FCS curves for such situation, and discuss how the information on this anomalous yet non-Gaussian diffusion can be extracted from these curves,
if at all. 

In our examples we mostly consider the two-dimensional situation (diffusion in a membrane), and use the dimensional units corresponding approximately
to the parameter of the FCS apparatus used in work \cite{us}. 

 
\section{Equilibrated CTRW}

The equilibrated CTRW is a nice exactly solvable model of a stationary random process whose displacements' PDF is not scaling as a function of elapsed time.

\subsection{Lower moments of displacements}
In a CTRW a random walker performs instantaneous jumps  choosing its displacement according to a probability density function $p(\textbf{r})$ and waiting time between 
two jumps according to a PDF of waiting times $\psi(t)$. For the ordinary process (starting at a step of the walker) the PDF of particles' displacement in Fourier-Laplace 
representation (i.e. the Laplace transform of the characteristic function of the displacements' distribution, $P(\textbf{k},s)=\int_0^\infty \langle e^ {-i\mathbf{kr}}\rangle e^{-st}dt$) 
obeys the Montroll-Weiss equation 
 \begin{equation}
  P(\textbf{k},s)= \frac{{1-\psi(s)}}{s} \frac{1}{1-\lambda(\textbf{k}) \psi(s)}.
 \label{pdf}
 \end{equation}
Here $\lambda(\textbf{k})$ is the characteristic function of displacements
in a single step being the Fourier transform of  $p(\textbf{r})$,
and $\psi(s)$ is the Laplace transform of waiting time PDF. For the equilibrated one one gets a slightly different form:
 \begin{equation}
    P_1(\textbf{k},s)= \frac{{1-\psi_1(s)}}{s}+ \frac{{1-\psi(s)}}{s} \psi_1 \frac{ \lambda(\textbf{k})}{1-\lambda(\textbf{k}) \psi(s)},
   \label{agedpdfg}
  \end{equation}
where $\psi_1(s)$ is the Laplace transform of the forward waiting time for the first step after the beginning of the observation \cite{book}. 
In our following discussions and simulations we will concentrate on a two dimensional case and on the Gaussian distribution of step length,
first, however, we will discuss the situation in some generality. 

For the waiting time distribution, however, we will stick to a particular model waiting time PDF, namely to a one sided L\'evy function with exponential cut off  (tempered stable law \cite{Rosin})
   \begin{equation}
 \psi(t) =  A \exp\left(-\frac{t}{t_c} \right)L_{\alpha}\left(\frac{t}{t_0}\right).
 \label{wtpdf}
 \end{equation}
Here $t_0$ is a characteristic time scale of a jump, $t_c$ is the cutoff time, and $A(t_0,t_c)$ is a normalization constant. 
In what follows we will put $t_0$ to unity (i.e. all times are now measured in units of $t_0$). In this case $A= \exp(t_c^{-\alpha})$. 
The Laplace transform of $\psi(t)$ then reads: 
  \begin{equation}
 \psi(s) =  \exp \left\{\left[-\left(s+\frac{1}{t_c} \right)^\alpha + \frac{1} {t_c^\alpha} \right]\right\}.
 \label{wtpdfs}
 \end{equation}
Exponential cut-off enforces the waiting time PDF to possess a mean which leads to establishing equilibrium at long times after process has started.
In the case the waiting time PDF possesses a mean, the forward waiting time PDF in Laplace domain has a form \cite{shlesinger1974asymptotic}
\begin{equation}
 \psi_1(s) =  \frac{1-\psi(s)}{s \tau}.
 \label{wtpdfg}
 \end{equation}
 with $\tau$ being the mean waiting time and can be calculated from first derivative of  waiting time PDF in Laplace domain 
\begin{equation}
\tau= - \left.\frac{\partial\psi(s) }{\partial s} \right|_{s = 0} = \alpha {t_c}^{\alpha-1}
 \label{mwtpdfaged}
 \end{equation}
 Now Eq.(\ref{agedpdfg}) reads:
 \begin{equation}
    P_1(\textbf{k},s)= \frac{{s\tau-1+\psi(s)}}{s^2 \tau}+ \frac{[1-\psi(s)]^2}{s^2 \tau}\frac{ \lambda(\textbf{k})}{1-\lambda(\textbf{k}) \psi(s)}.
   \label{agedpdfg1}
  \end{equation}

 Let us now turn to the spatial part of the distribution. 
Since the characteristic function of the displacement distribution is the generating function of its moments, the Taylor expansion of the 
this function gives in the isotropic $d$-dimensional situation for the two lowest non-vanishing moments, namely for the second one, $m_2$ and for the fourth one $m_4$ 
\cite{franosch}:
\begin{equation}
p(\mathbf{k})=1-\frac{k^2}{2d} m_2 + \frac{k^4}{8d(d+2)}m_4 +O(k^6). 
\label{moment_exp}
\end{equation}
Substituting 
\[
\lambda(\mathbf{k}) = 1-\frac{k^2}{2d}  \langle l^2 \rangle + \frac{k^4}{8d(d+2)}  \langle l^4 \rangle +O(k^6) 
\]
(with $ \langle l^2 \rangle$ and $ \langle l^4 \rangle$ being the second and the fourth moment of the step length, respectively) into Eq.(\ref{agedpdfg1}) 
and expanding this in Taylor series in $k$ we get:
\begin{eqnarray*}
&& P_1(\mathbf{k},s) =  \frac{1}{s} - \frac{k^2}{2d} \frac{ \langle l^2 \rangle}{s^2 \tau} \\
&&+ \left[\frac{1}{4d^2} \frac{\psi(s)}{1-\psi(s)} 
\frac{1}{s^2 \tau}  \langle l^2 \rangle^2 + \frac{1}{8d(d+2)}\frac{1}{s^2 \tau}  \langle l^4 \rangle \right] k^4 + O(k^6).
\end{eqnarray*}
Comparing this with the moment expansion, Eq.(\ref{moment_exp}), which now reads
\begin{equation} 
   P_1(\textbf{k},s)=1-\frac{k^2}{2 d} \delta r^2(s) + 
   \\\frac{k^4}{8 d (d+2)} \delta r^4(s) + O(k^6)
 \label{dd}
  \end{equation}
where $\delta r^2(s)$ and $\delta r^4(s)$ are the second and the fourth moments of the displacements' PDF in the Laplace domain,
we get that
\begin{equation}
\delta r^2(t)= \frac{\langle l^2 \rangle}{s^2 \tau}.
\label{secmomages}
\end{equation} 
The second moment of PDF of distribution has a generic form and is independent of the particular form of the waiting time PDF \cite{tunaley1975}.    
Its inverse Laplace transform grows linearly with time:
\begin{equation}
\delta r^2(t)= \frac{\langle l^2 \rangle}{\tau} t.
\label{secmomaget}
\end{equation}
The fourth moments is
\begin{equation}
\delta r^4(s)  = \frac{ \langle l^4 \rangle}{s^2 \tau} +  \frac{2(d+2)}{d} \frac{1}{s^2 \tau} \frac{\psi(s)}{1-\psi(s)} \langle l^2 \rangle^2
\label{fourthfs}
\end{equation}
in the Laplace domain. Translating this back to the time domain gives different results for long times $t \gg t_c$ and 
for short times $1 \ll t \ll t_c$, corresponding to $s \ll 1/t_c$ and to $1/t_c \ll s \ll 1$ 
respectively (note that the non-universal domain $s \ll 1$ is of no interest here). In the long time domain $\psi \approx 1 - s \tau$ and therefore
\[
\delta r^4(s)  = \frac{ \langle l^4 \rangle}{s^2 \tau} +  \frac{2(d+2)}{d} \frac{1}{s^3 \tau^2} \langle l^2 \rangle^2
\]
which translates into
\begin{equation}
\delta r^4(t) = \langle l^4 \rangle\frac{t }{\tau}  + \frac{(d+2)}{d} \langle l^2 \rangle^2 \frac{t^2}{\tau^2}
\label{fourthflass}
\end{equation}
and is dominated by the second term.  For $1 \ll t \ll t_c$ we have $\psi(s) \approx 1 -s^\alpha$,
\[
 \delta r^4(s)=  \frac{ \langle l^4 \rangle}{s^2 \tau} +  \frac{2(d+2)}{d} \frac{1}{s^{2+\alpha} \tau} \langle l^2 \rangle ^2
\]
which in the time domain translates into
\begin{equation}
 \delta r^4(t)=  \langle l^4 \rangle\frac{t }{\tau}   +  \frac{2(d+2)}{d\Gamma(2+\alpha)} \langle l^2 \rangle^2 \frac{t^{1+\alpha}}{\tau}
 \label{4long}
\end{equation}
indicating strong non-Gaussianity of PFD of displacement in short times, and the absence of scaling of the distribution as a whole.


\subsection{The form of the PDF}

The Montroll-Weiss approach is a very nice one to get the moments of the distribution, but is less convenient for obtaining the 
form of the PDF in the space-time domain, since it involves both the inverse Laplace and the inverse Fourier transforms which have to be performed numerically.
One can save on the Fourier transform when applying the subordination approach \cite{book}. The same approach will be used for direct calculation of the 
FCS autocorrelation curves, and therefore is discussed here in some detail.
The PDF of displacement can be written as  
\begin{equation}
P_1(\mathbf{r},t) = \sum_{n=0}^{\infty} \chi_n(t) P(\mathbf{r},n)
\label{gg}
\end{equation}
where $P(\mathbf{r},n)$ is the PDF of particle's displacements in a simple random walk after $n$ steps, and $\chi_n(t)$ being the probability of taking exactly $n$ steps up to the time $t$. 
The number of steps is the internal temporal variable of the subordination scheme (operational time).

For an aged or equilibrated CTRW the probability $\chi_0(t)$ of making no step from the starting point of measurement $t=0$ till time $t$
is given in the Laplace domain by 
\[
 \chi_0(s)=\frac{1-\psi_1(s)}{s},
\]
and all other $\chi_n(s)$ are
\[
 \chi_n(s)= \psi_1(s) [\psi(s)]^{n-1}  \frac{1-\psi(s)}{s}.
\]
In Laplace domain Eq.(\ref{gg}) transforms to
\begin{eqnarray}
P_1(\mathbf{r},s) =&& \frac{1-\psi_1(s)}{s} \delta(r) \nonumber \\ 
&&+  \psi_1(s)   \frac{1-\psi(s)}{s}  \sum_{n=1}^{\infty} [\psi(s)]^{n-1} P(\mathbf{r},n).
\label{sub1}
\end{eqnarray}
For $n$ large the function $P(\mathbf{r},n)$ can be well approximated by a Gaussian
\begin{equation}
 P(\mathbf{r},n)= \frac{1}{\pi \langle l^2 \rangle n} \exp \left( - \frac{r^2}{ \langle l^2 \rangle n} \right). 
 \label{Gauss}
\end{equation}
Eq.(\ref{sub1}) is then rewritten as follows: 
Each term in the sum is multiplied by $\psi(s)$, and the whole sum is then divided by the same function, and $\psi_1(s)$ is written in the explicit form given by 
Eq.(\ref{wtpdfg}). Now we obtain: 
\begin{eqnarray}
 P_1(\mathbf{r},s) =&& \left[ \frac{1}{s}- \frac{1-\psi(s)}{s^2 \tau}\right] \delta(r)  \\
 &&+      \frac{[1-\psi(s)]^2}{s^2 \tau \psi(s)} \sum_{n=1}^{\infty} [\psi(s)]^n P(\mathbf{r},n). \nonumber 
\label{sub2}
\end{eqnarray}
For $t \gg t_0$ 
the typical number of steps taken is large and $n$ can be considered as a continuous variable, and the sum changed to the integral:
\begin{eqnarray}
  P_1(\mathbf{r},s) =&& \left[ \frac{1}{s}- \frac{1-\psi(s)}{s^2 \tau}\right] \delta(r)  \\
  && + \frac{[1-\psi(s)]^2}{s^2 \tau \psi(s)} \int_0^\infty P(\mathbf{r},n) e^{n \ln \psi(s)} dn \nonumber
  \label{subf1}
\end{eqnarray}
The integration corresponds to the Laplace transform of $P(\mathbf{r},n)$ in its temporal variable $n$: 
\[
\int_0^\infty P(\mathbf{r},n) e^{n \ln \psi(s)} dn=\tilde{P}[\mathbf{r},-\ln \psi(s)].
\]
Here and below the tilde denotes the Laplace transformed, when it is not evident from the variable used as the Laplace frequency, i.e. when it is not simply $s$. 
We note that typically one expands $-\ln \psi(s)$, but for the PDF given by Eq.(\ref{wtpdf}) this logarithm is simpler than its expansion! Thus, 
\begin{eqnarray}
P_1(\mathbf{r},s) =&& \left[ \frac{1}{s}- \frac{1-\psi(s)}{s^2 \tau}\right] \delta(r)  \nonumber \\  
&&+ \frac{[1-\psi(s)]^2}{s^2 \tau \psi(s)} \tilde{P}[\mathbf{r},-\ln \psi(s)].
 \label{subf}
\end{eqnarray}
Taking the Laplace transform of  $P(\mathbf{r},n)$, Eq.(\ref{Gauss}), explicitly, we get 
\begin{eqnarray}
   P_1(\mathbf{r},s) = &&\left[ \frac{1}{s}- \frac{1-\psi(s)}{s^2 \tau}\right] \delta(r)  \\
 && + \frac{[1-\psi(s)]^2}{s^2 \psi(s) \tau } \frac{2}{\pi \langle l^2 \rangle}  K_0 \left[2 r \sqrt{\frac{{-\ln \psi(s)}}{\langle l^2 \rangle}} \right]  \nonumber 
 \label{subpdf}
 \end{eqnarray}
with $K_0(z)$ being the modified Bessel function of the second kind.   
 \begin{figure}[h!!]
\begin{center}
\includegraphics[width=0.5\textwidth]{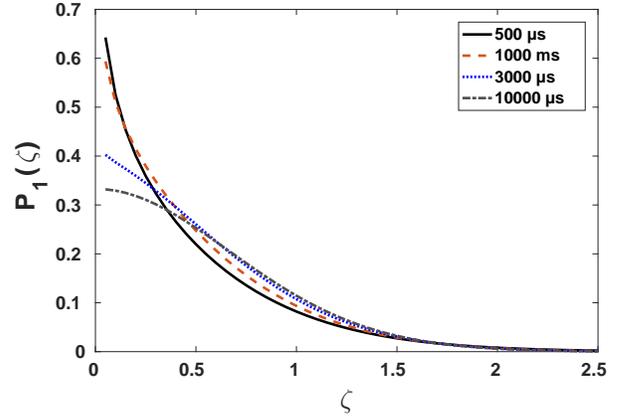}
\caption{The probability density function of displacements as obtained by numerical inversion of Eq.\ref{subpdf} for times $t$ being $500$, $1000$, $3000$ and $10000$ $\mu s$ with $t_c=1000$ $\mu s$ and $\alpha=0.6$.
See text for details.}
\label{pdfp}
\end{center}
\end{figure}
The numerical inversion of this expression can be easily performed using the Gaver-Stehfest algorithm \cite{stehfest}. The results for different times are
shown in Fig. \ref{pdfp}. To compare the forms of the PDF for different times, 
we plot them as a function of a dimensionless variable $\zeta =\frac{r}{\sqrt{{\delta r}^2 (t)}}$, which is the displacement normalized with respect to r.m.s. displacement.   
Fig.~\ref{pdfp} shows the PDFs  for times $t$ being $500$, $1000$, $3000$ and $10000$ $\mu s$ with $t_c=1000$ $\mu s$ and $\alpha=0.6$ (the rapidly decaying $\delta$-peak at zero is not shown). 

\begin{figure}[h!!]
\begin{center}
\includegraphics[scale=.5]{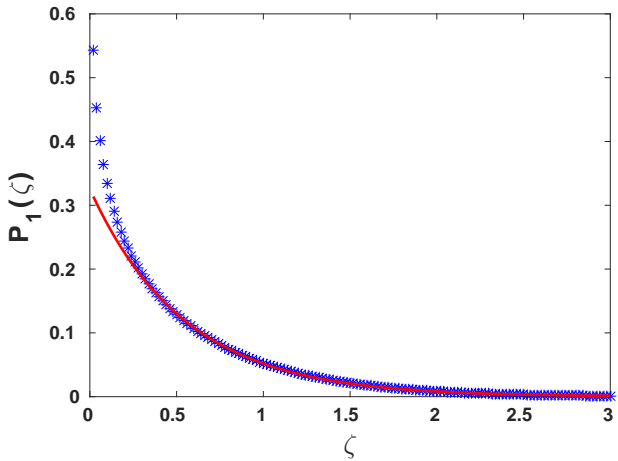}
\caption{Probability density function of displacement for  $t=500$  $\mu s $ (symbols). Red line shows the exponential fit to the tail of the distribution ($\zeta > 0.5$) plotted in the whole $\zeta$-domain.}
\label{expfits}
\end{center}
\end{figure}

The form of the distribution for $t < t_c$ (here $t=500$ $\mu s$) is shown in Fig.~\ref{expfits}. 
Analyzing the distribution at short times indicates that they possess an exponential tail (resembling the prediction of the diffusing diffusivity model) by higher peak. 
Similar behaviour has been found in many experimental works, mentioned in introduction, using sampling the PDF from SPT measurements.  

As it could be concluded from Fig.~\ref{pdfp}, this distribution for short times is strongly peaked, and tends to a Gaussian at $t \gg t_c$.

\begin{figure}[h!!]
\begin{center}
\includegraphics[scale=0.48]{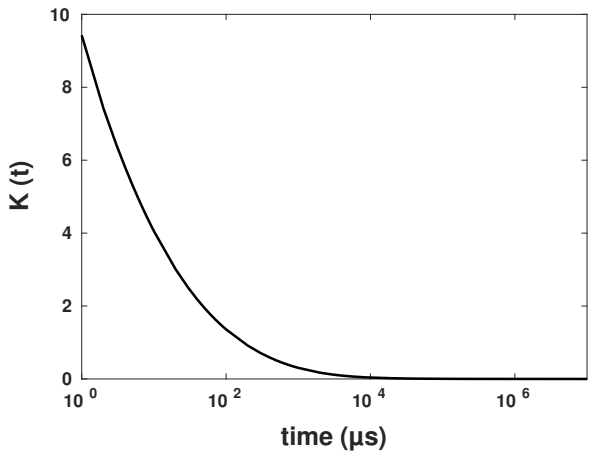}
\caption{Reduced kurtosis, Eq.\ref{rmK}, as calculated using the previous result and Eq.(\ref{secmomaget}) for $t_c=1000$  $\mu s $ and $\alpha=0.6$. 
Deviations from zero indicate strong non-Gaussianity; these vanish for vanishes for $t \gg t_c$. }
\label{rkk}
\end{center}
\end{figure}

The deviations from the Gaussian can be easily characterized when considering the reduced kurtosis 
  \begin{equation}
     K= \frac{d}{d+2} \frac{\delta r^4 (t)}{\delta r^2 (t)}-1,
     \label{rmK}
  \end{equation}
as following from Eqs.(\ref{secmomages}) and Eq.(\ref{fourthflass}) or Eq.(\ref{4long}). At times $1 < t < t_c$ the second term in Eq.(\ref{4long}) dominates unless the
fourth moment in a single step displacement is exceedingly large. The distribution then is highly leptocurtic, and the reduced kurtosis decays essentially as $t^{\alpha-1}$.
At long times, when Eq.(\ref{fourthflass}) is applicable and the result is dominated by its second term. The reduced kurtosis vanishes, indicating the 
transition to a Gaussian. This behaviour is clearly seen in Fig.\ref{rkk} which shows the reduced kurtosis as a function of time as obtained by numerical Laplace inversion of Eq.(\ref{fourthfs}):
Extremely strong deviations from Gaussianity at short times decay at times larger than the characteristic cut-off time $t_c$.

\section{Fluorescence correlation spectroscopy}
  
Let us now discuss the consequences of the absence of scaling in PDF for the fluorescence correlation spectroscopy. 
In FCS labeled tracers enter and exit a detection volume, defined by a focal volume of a laser beam, and emit fluorescence photons upon the  excitation.  
The observed photon counts are translated into the intensity-intensity correlation function $G(t)$ which is the main outcome of the experiment.

Generally FCS curve can be expressed via a single integral of the PDF of the absolute displacement, $P(r,t)$, and an apparatus function, $F(r)$, 
defining the laser intensity distribution in the detection volume \cite{us}: 
\begin{equation}
G(t)= \int_0^\infty P(r,t) F(r) dr.
\label{fcsgen}
\end{equation}  
Assuming a Gaussian intensity distribution in the detection volume, the apparatus function in two dimensions reads
\begin{equation}
F(r) = 2 \pi r \exp\left(-\frac{r^2}{r_0^2}\right). 
\label{General2}
\end{equation}
with $r_0$ being the beam's waist.
%
%
%
We note that $G(t)$, given by Eq.(\ref{fcsgen}), is a linear functional of $P(r,t)$ involving only the manipulation of spatial variables 
(weighted integration). Therefore, one can again use the subordination approach.
Thus first pass to the Laplace domain, $G(s)$ via $P(r,s)$, and then perform the the numerical inversion to the time domain.

\begin{figure}[t!]
\begin{center}
\scalebox{0.5}{\includegraphics{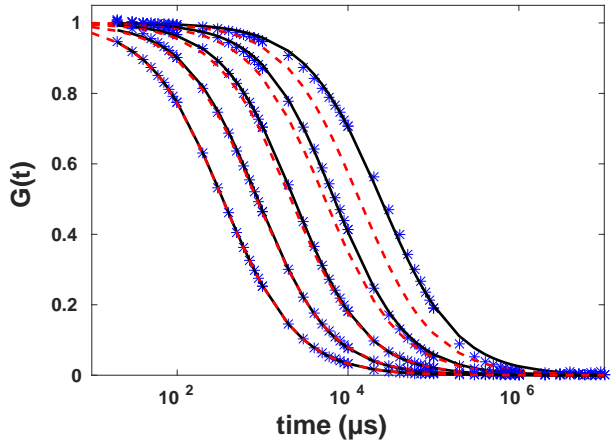}}
\caption{The FCS curves obtained by numerical Laplace inversion of Eq.(\ref{finalfcs}) (symbols) together with the FCS curves for normal diffusion with the corresponding diffusion coefficient (dashed lines)
and their best fits to Eq.(\ref{standard}) (full lines).
The curves correspond to $t_c$ being $10$, $100$, $1000$,$10000$ and $100000$ $\mu s$ from left to right.}
\label{tfcs}
\end{center}
\end{figure}

Let us calculate
\[
 G_1(s) = \int_0^\infty P_1(r,s) F(r) dr.
\]
Substituting the series for $P_1(r,s)$ given by Eq.(\ref{subf1}) and  performing term-by-term integration over spatial variables leads  to the expression
\begin{eqnarray}
   G_1(s) = && \frac{1}{s}- \frac{1-\psi(s)}{s^2 \tau} \\ 
   && + \frac{[1-\psi(s)]^2}{s^2 \tau \psi(s)}  \sum_{n=0}^{\infty} [\psi(s)]^n G(n)  \nonumber
\label{fcsctrwap}
 \end{eqnarray}
with $G(n)$ coinciding with the FCS curve for normal diffusion in the operational time:
\[
 G(n)= \left(\frac{ \langle l^2 \rangle}{r_0^2} n +1\right)^{-1}.
\]
Similar to the previous discussion, the summation corresponds to the Laplace transform of $G(n)$ in the changed Laplace variable:
\[
 \sum_{n=0}^{\infty} [\psi(s)]^n G(n) \approx \int_0^\infty G(n) e^{n \ln \psi(s)} dn=\tilde{G}[-\ln \psi(s)].
\]
Replacing the explicit form of the Laplace transform of  $G(n)$ we get 
\begin{eqnarray}
   G_1(s) = && \frac{1}{s}- \frac{1-\psi(s)}{s^2 \tau}+\frac{[1-\psi(s)]^2}{s^2 \psi(s) \tau } \frac{r_0 ^2 }{\langle l^2 \rangle} \times \label{finalfcs} \\
 &&\exp\left[ -\frac{r_0 ^2 \ln \psi(s)}{ \langle l^2 \rangle} \right] \Gamma\left[0, - \frac{r_0 ^2 \ln \psi(s)}{\langle l^2 \rangle } \right] \nonumber 
 \end{eqnarray}
with $\Gamma(a,z)$ being an incomplete Gamma function. 
The results of numerical inversion of this expression are shown in Fig.~\ref{tfcs}. 
Here we plot the results of numerical inversion of Eq.(\ref{finalfcs}) with the same Gaver-Stehfest algorithm (symbols)
for $t_c$ being $10$, $100$, $1000$, $10000$ and $100000$ $\mu s$. The numerical FCS data points were fitted to a standard
two-parametric expression
\begin{equation}
  G(t) = \left(\frac{4 D_\beta}{r_0^2} t^\beta +1\right)^{-1}
  \label{standard}
 \end{equation} 
with free parameters $\beta$ and $D_\beta$; the results of these fits are shown as full lines. 
In addition we plot the FCS curves which would be obtained for the case of normal diffusion 
\begin{equation}
  G(t) = \left(\frac{4 D}{r_0^2} t +1\right)^{-1}
  \label{normaldiff}
\end{equation} 
with the terminal diffusion coefficient $D = \langle l^2 \rangle /4 \tau$.  
While for small $t_c$ the differences between the inversion results and their fits from the normal diffusion curves is minor, they gets larger as $t_c$ increases. 
At larger $t_c$s Fitting numerical FCS data to Eq.(\ref{standard}) still indicates normal diffusion with $\beta = 1 \pm 0.03$,
but with the diffusion coefficient $D_1$ which is considerably smaller than the terminal diffusion coefficient $D$.    
For example, for the largest $t_c=10^5 \; \mu s$ the terminal diffusion coefficient is $D = 0.0041$ while the one obtained by the fit is almost two times
smaller: $D_{\mathrm{fit}} = 0.0023$.

\begin{figure}[t!]
\begin{center}
\scalebox{0.5}{\includegraphics{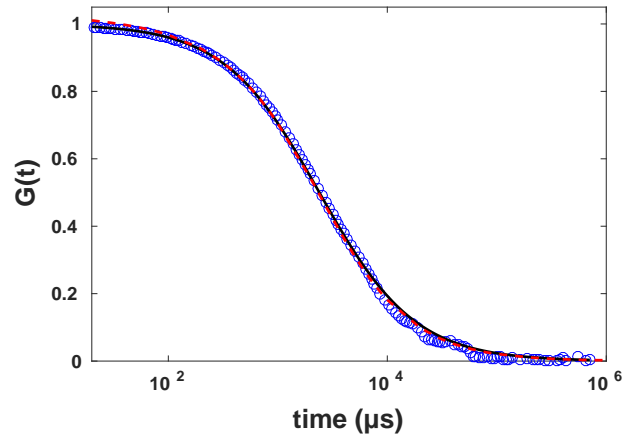}}
\caption{Simulated FCS curve (open circles) together with its fit to Eq.\ref{standard} (full line, color online: black), and with the result of the numerical inversion of Eq.(\ref{finalfcs})
(dashed line, color online: red). }
\label{fcss}
\end{center}
\end{figure}

To check our theoretical predictions we also performed an exemplary lattice simulation of CTRW for the values of parameters used in our calculations. 
Simulations are done using the standard random walk on a square lattice with equal probability to jump to four nearest neighbors.  
Having arrived to a site particles were set to wait for a time chosen from waiting time according to Eq.\ref{wtpdf} with $t_0 = 1 \mu s$ and  $t_c=1000 \mu s$. 
The waiting times were generated using the acceptance-rejection method \cite{accrej}: First a random variable $y$ distributed according to a  L\'evy law is generated. Then a random variable $u$ 
uniformly distributed between $0$ and $1$ was generated. Waiting times meeting the condition, $\exp(-y/t_c)<u $, were accepted as actual waiting times, otherwise the trial was repeated. 
The size of the simulation box was taken to be  $L=300$ in each direction, with  periodic boundary conditions. The lattice constant representing the step length was taken to be
$1 \mu m$. The confocal volume with $r_0=15 \mu m$ was positioned in the center of box and intensities according to the distance of particles from the center of confocal volume were calculated at each step of
simulation, $I_i=\exp(-r^2/r_0^2)$. The total number of particles was chosen such that their mean number within the confocal volume did not exceed unity. Simulation is run for $10^8$ time steps. 
The results for these values of parameters are plotted in Fig.~\ref{fcss}.  This parameters chosen correspond to the middle curve (third from the left) from the previous Fig. \ref{tfcs}. 
This exemplary result shows the applicability of our semianalytical approach. The deviations at very short times are mainly caused by the lattice nature of the model used in simulations. 

\section{Asymptotic behavior}

Let us now investigate the short and long time asymptotic of Eq.(\ref{finalfcs}) with respect to the characteristic time $t_c$.
For $t \gg t_c$  ($s \ll 1/t_c$),  $\psi(s)$ can be approximated by $\psi(s) \approx 1-\tau s$ and $-\ln \psi(s) \approx \tau s$. 
Substituting this approximation into Eq.(\ref{finalfcs}) we see that the first two terms vanish, and the third term coincides with the Laplace transform of $G(n)$ with $n=t/ \tau$. 
Therefore in this asymptotic domain  
\begin{equation}
 G_1(t) \sim  \left(\frac{\langle l^2 \rangle}{\tau r_0^2} t +1\right)^{-1} 
 \label{longasym}
\end{equation}
which is the normal diffusion with terminal diffusion coefficient $D=\langle l^2 \rangle /4 \tau$. 

\begin{figure}[h!!]
\begin{center}
\includegraphics[width=0.5\textwidth]{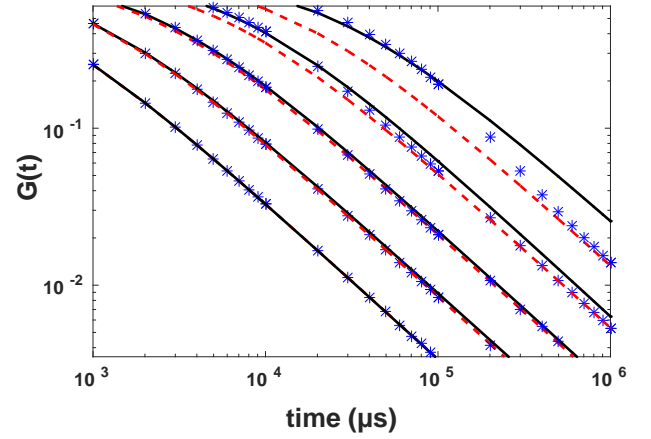}
\caption{The FCS results from Eq.\ref{finalfcs} (stars) together with the FCS curves for normal diffusion with
the terminal diffusion coefficient (dashed lines) and the fitted curves from Fig.~\ref{tfcs} (full lines) at long times. The transition between the intermediate domain
and the terminal asymptotic behavior is clearly see for larger $t_c$. The curves correspond to $t_c$ being $10$, $100$, $1000$,$10000$ and $100000$ $\mu s$ from left to right. 
Note the double logarithmic scales.}
\label{tfcssl}
\end{center}
\end{figure}

\begin{figure}[h!!]
\begin{center}
\includegraphics[width=0.5\textwidth]{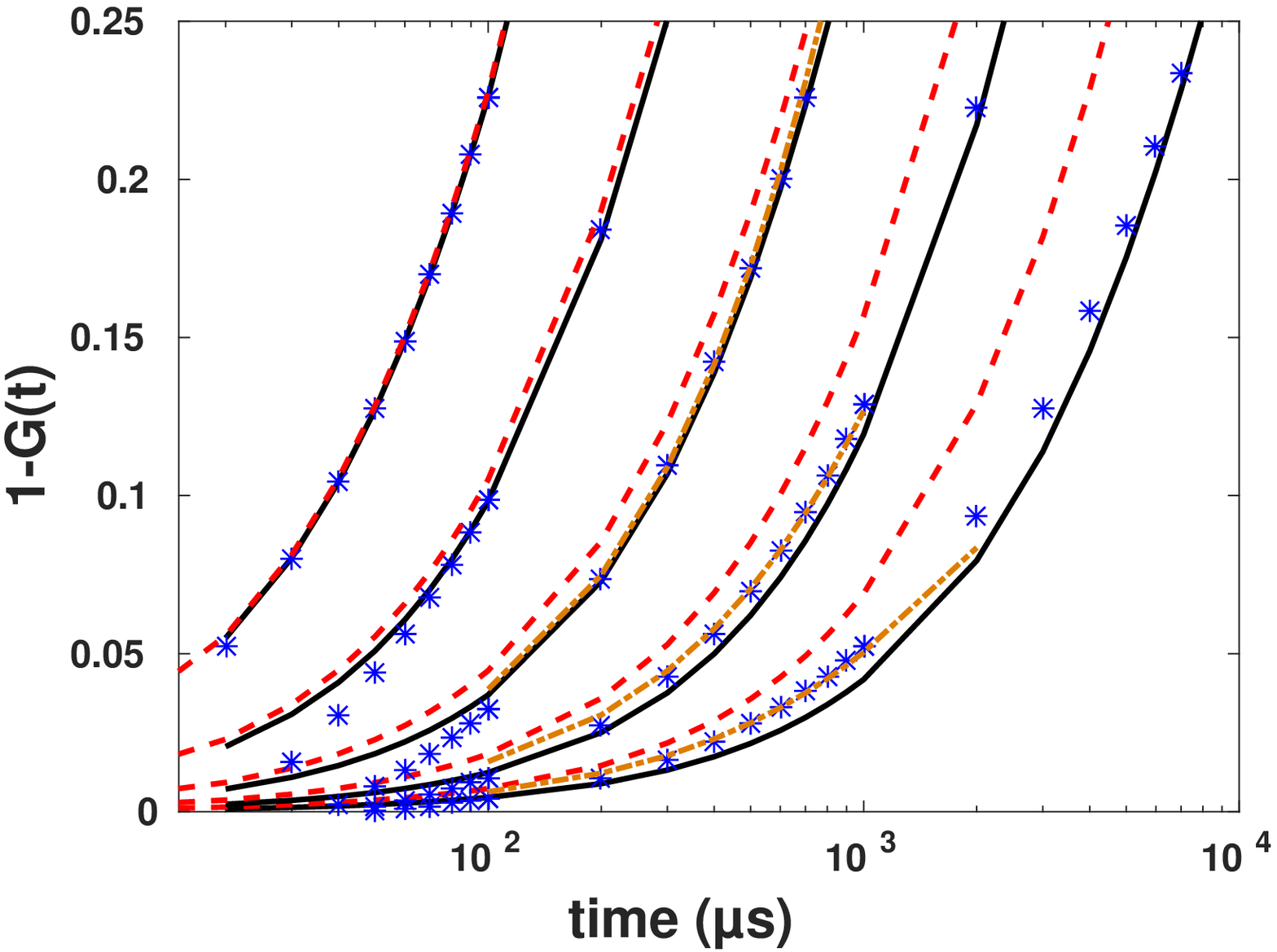}
\caption{The FCS results from Eq.\ref{finalfcs} (stars) together with the FCS curves for normal diffusion with
the terminal diffusion coefficient (dashed lines) and the fitted curves from Fig.~\ref{tfcs} (full lines) at short times.
The curves correspond to $t_c$ being $10$, $100$, $1000$,$10000$ and $100000$ $\mu s$ from left to right.
Curves corresponding to the theoretical approximation for the short times from Eq.\ref{shorty} are plotted 
as dotted lines for the time domain where the approximation is applicable.}
\label{tfcss}
\end{center}
\end{figure}

For short times, $t_0\ll t \ll t_c$  ($1\gg s \gg  1/t_c$), one has  
$\psi(s) \approx 1-s^\alpha $ and $\ln \psi(s) \approx -s^\alpha$, and therefore Eq.\ref{fcsctrwap} reads as
\[
   G_1(s)\approx \frac{1}{s} - \frac{s^{\alpha -2}}{\tau}+ \frac{s^{2\alpha -2}}{\tau } \int_0^\infty  G(n) e^{-n s^\alpha} dn.
\]
Expanding $G(n)$ as
\[
   G(n)\approx 1+ \frac{\langle l^2 \rangle}{{r_0}^2} n+{\frac{{\langle l^2 \rangle}^2}{2{r_0}^4}} n^2
\]
and performing term-by-term integration one gets   
\[
   G_1(s)\approx \frac{1}{s} - \frac{ \langle l^2\rangle}{\tau {r_0}^2} s^{-2} +{\frac{{\langle l^2 \rangle}^2}{{\tau }{r_0}^4}} s^{-\alpha -2}.
\]
The inverse Laplace transform of the expression above reads: 
\begin{equation}
   G_1(t)\approx 1- \frac{ \langle l^2\rangle}{\tau {r_0}^2} t +{\frac{{\langle l^2 \rangle}^2}{{\tau }{r_0}^4}} \frac{t^{1+\alpha }}{\Gamma(1+\alpha)}.
   \label{shorty}
\end{equation}
The same results for the short time asymptotic of FCS curve can be derived immediately from Eq.(\ref{fcsgen}) using the approach of  \cite{us}. Expanding the apparatus function one gets
\[
G(t)=1- \frac{\delta r^2(t)}{r_0 ^2} + \frac{\delta r^4(t)}{2 r_0 ^4}+... \;.
\]
Using the explicit expressions for the moments and noting the the terms containing $\langle l^2\rangle^2$ are negligible in the time window considered, namely for $t \gg t_0$,
we again arrive at Eq.(\ref{shorty}). The corresponding behaviors are analyzed in Figs.~\ref{tfcssl} and \ref{tfcss}.
Fig.~\ref{tfcssl} indicates the asymptotic transition of FCS data to Eq.(\ref{longasym}). This asymptotic behavior is clearly seen in the last two set of data 
corresponding to $t_c= 10000 \mu s$ and $100000 \mu s$. 

Expected behaviour for the short time asymptotic from Eq.(\ref{shorty}) is also
observed in Fig.~\ref{tfcss}. The approximations we used to derive this equation are actually applicable for the large $t_c$ in this figure, 
for the time window $t \ll t_c  $ and $t \gg \tau $.

Regardless of exact behavior of FCS curve for very short times, the deviation of data from the normal diffusion curves in the short and long time 
domain is the only way to spot normal, yet non-Gaussian behaviour. In experiment this deviations might be obscured by noise being strong
exactly in these domains, and the short-time behaviour might be strongly influenced by the photophysical 
properties of the dye. The conclusion is that the strongly non-Gaussian behaviour at short times, and the absence of scaling of the displacements' PDF as a whole
may stay unnoticed in the FCS experiments, which will hint onto normal diffusion, but deliver a wrong estimate for the diffusion coefficient. 

\section{Summary and Conclusions}
In summary, we considered  the CTRW of tracers with waiting time probability density function following a L\'evy-stable law with  exponential cut off (tempered stable law) in equilibrium in two dimensions.
This is the case corresponding to the essentially normal diffusion, with the second moment of displacement growing linearly in time. The PDF of particles' displacements is however 
strongly non-Gaussian at shorter times, and does not scale. The kurtosis and the form of the corresponding PDF are obtained semi-analytically by means of numerical Laplace
inversion of the corresponding expressions in the Laplace domain. The equilibrated CTRW with tempered stable waiting time density is therefore a nice theoretical model
for investigating the effects of the absence of scaling on different physical properties of systems governed by diffusion and on the outcomes of the corresponding
experiments. In the present work we concentrate on the possible outcomes in fluorescence correlation spectroscopy being a method of choice in investigating 
diffusion processes in living cells and in crowded in-vitro systems. The results show that although the deviations from Gaussian behavior may be detected when
analyzing the short- and long-time asymptotic behavior of the corresponding curves (which experimentally are always obscured by noise and by photophysical effects),
the bodies of the curves are still perfectrly fitted by the fit forms obtained for normal diffusion. The diffusion coefficients obtaied from the fits may however differ 
considerably from the true tracer diffusion coefficient as describing the time-dependence of the mean squared displacement of the tracer. 
 
\bibliography{sample1}

\end{document}